\documentclass[a4paper,11pt]{article}
\usepackage{pos}

\begin{document}

\title{The Correlated Spatial Structure of the Proton: Two-body densities as a framework for dynamical imaging}

\author*[a]{Zaki Panjsheeri}
\author[a]{Joshua Bautista}
\author[a]{Simonetta Liuti}

\affiliation[a]{University of Virginia Physics Department,\\
  382 McCormick Rd, Charlottesville, Virginia}

\emailAdd{zap2nd@virginia.edu}
\emailAdd{rzp9zf@virginia.edu}
\emailAdd{sly4@virginia.edu}

\abstract{We present results on two-parton densities in coordinate space 
which capture a fuller dynamical picture of the proton’s internal structure, including information on the relative position between quarks and gluons in the transverse plane. The connection of such two-body densities to observables, proceeds in QCD, via the definition of  double generalized parton distributions (DGPDs).}

\FullConference{25th International Spin Physics Symposium (SPIN 2023)\\
 24-29 September 2023\\
 Durham, NC, USA\\}


\maketitle

\section{Introduction}
It is believed that the exclusive process of deeply virtual Compton scattering (DVCS) allows us to access generalized parton distributions (GPDs) \cite{Ji:1996ek,Ji:1996nm,Radyushkin:1997ki} (we refer the reader to Refs.\cite{Diehl:2001pm,Belitsky:2005qn,Kumericki:2016ehc} for reviews on the subject). In turn, GPDs, through Fourier transformation, give spatial information on the charge and matter distributions of the quarks and gluons inside the nucleon \cite{Burkardt:2000za}. 
%
The physical properties derived from the Fourier transforms of GPDs include the spatial distributions of each partonic component of the nucleon in the transverse plane with respect to the proton's motion, as well as the orbital component of angular momentum. 
Notwithstanding the wealth of information provided by these quantities, to capture a fuller dynamical picture of the proton’s internal structure, information on the {\it relative} position between partons is crucial. 
In order to access information on the relative position of quarks and gluons, one needs to define the correlation functions yielding the two-body density distribution in the transverse plane. 
Connecting the two-body densities to observables, we found that the latter can be defined in QCD with generalized double parton distributions (GDPDs). Using GDPDs, we can address additional observables to describe quark and gluon dynamics, including their overlap probabilities. 
Such quantities can be extracted from experimental measurements of deeply virtual exclusive processes characterized by multi-particle final states.

\section{From one body to two body density}
\begin{figure*}[h]
    \centering
    \includegraphics[width=4cm]{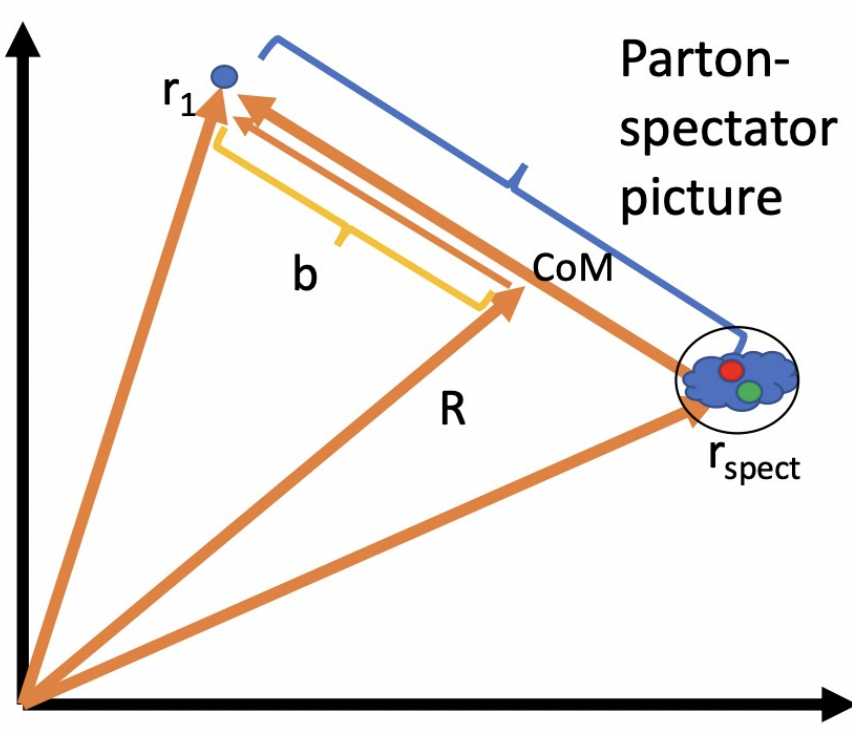}
    \hspace{2cm}
    \includegraphics[width=4cm]{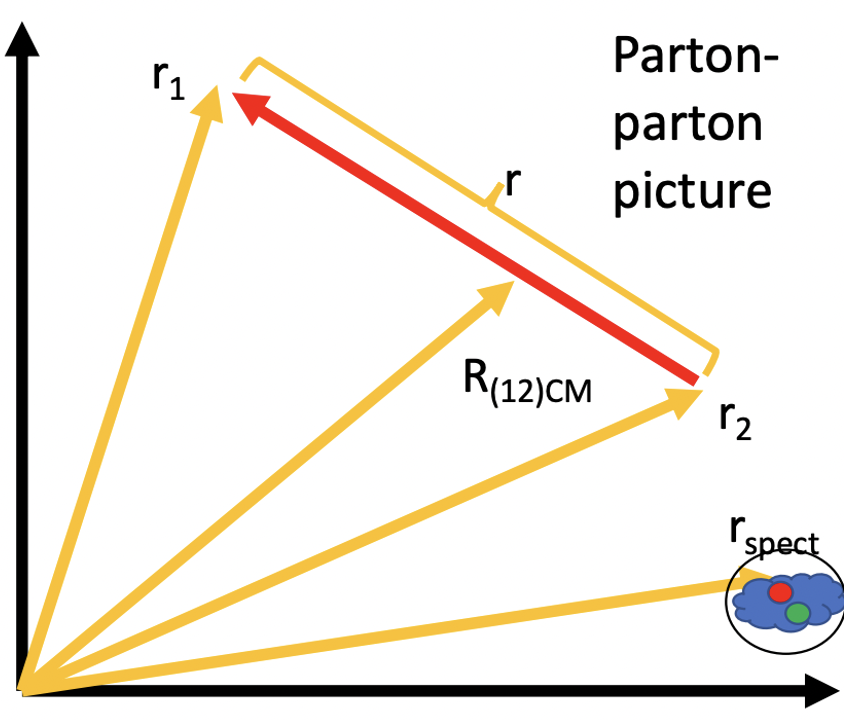}
    \caption{Spatial coordinates in the transverse plane for one-body distributions (left) and two-body distributions (right).}
    \label{fig:space_coord}
\end{figure*}

We focus our description on the unpolarized quark inside an unpolarized nucleon described by the GPD, $H^{q}(X, \zeta = 0, t)$, where $x$ is the quark longitudinal momentum fraction, $t = \Delta^2$ is the four-momentum transfer squared between the initial and final proton, and $\zeta$, the skewness parameter is set to zero. Through Fourier transformation, one obtains the impact parameter-dependent parton distribution function (IPPDF) $\rho(X,{\bf b}_{T})$, in terms of ${\bf b}_{T}$, the Fourier conjugate to the transverse momentum transfer, ${\bf \Delta}_T$ \cite{Burkardt:2000za, Diehl:2002he}. The spatial variables relevant for this problem are described in the left diagram in Figure \ref{fig:space_coord}, giving a partonic picture in coordinate space, as first observed using Light Cone (LC) kinematics in Ref.\cite{Soper:1976jc}. 

\begin{figure*}[h]
    \centering
   \includegraphics[width=7cm]{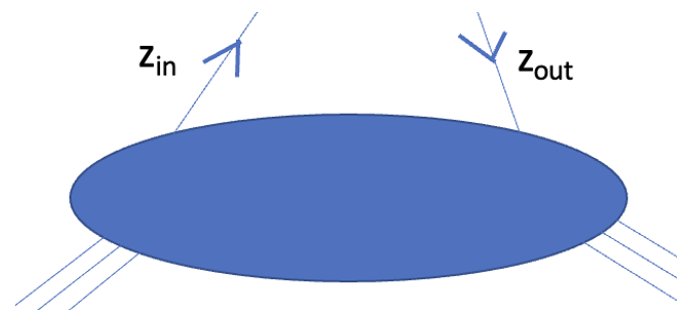}
    \caption{Correlation function for the GPD in momentum space}
    \label{fig:one_body_corr}
\end{figure*}

From the figure, one can see that two relevant variables describe the spatial configuration in a parton-spectator system, namely, $b=b_T$, the relevant transverse distance of the parton from the Center of Momentum (CoM), and $R$, the position of the nucleon's CoM.
To obtain such a picture through GPDs, we take the correlation function represented in momentum space by Fig. \ref{fig:one_body_corr}, 
\begin{equation}
\label{eq:1correlator}
W_{\Lambda,\Lambda'}^{\Gamma} = \int \frac{dz_{in}^{-}}{(2\pi)}   \int \frac{dz_{out}^{-} }{(2\pi)^2}   \, e^{i  (k_{in} z_{in})} \, e^{-i  (k_{out} z_{out})}  \: 
\langle p',\Lambda' | \overline{\psi} (z_{out}) \, \Gamma \, \mathcal{U}(z_{in},z_{out})\psi(z_{in})|p,\Lambda \rangle \Big |_{\substack{z_{in(out)}^{+} =0 \\ {\bf z}_{T,in(out)}=0} }, 
\end{equation}
We represent the four-vectors in LC notation, $v \equiv (v^+,v^-,{\bf v}_T)$, $v^\pm = (v^o \pm v^3)/\sqrt{2}$, $d^4 z = d z^- d z^+ d^2 {\bf z}_T$; we evaluate the fields at equal LC time, $z^+=0$. $\psi(z_{in})$, $\bar{\psi}_{out}(z)$ are the quark fields; $\Gamma$ is an operator between quark fields,  {\it i.e.} either a specific gamma matrix, or a combination of gamma matrices; finally, ${\cal U}(z_{in},z_{out})$, is the link ensuring  gauge invariance. In what follows we choose the LC gauge for which ${\cal U}(z_{in},z_{out})$ can be taken equal to one. The field operators product in Eq.\eqref{eq:1correlator} is evaluated between an incoming proton state, $\mid p, \Lambda \rangle$, of definite momentum, $p$, and helicity, $\Lambda$, and an outgoing proton state, $\mid p', \Lambda' \rangle$, with momentum, $p'=p-\Delta$, and helicity, $\Lambda'$.
%
%
Taking $\Gamma= \gamma^+$, and Fourier transforming, one obtains
\footnote{The correlation function's parametrization in terms of GPDs also depends on the QCD evolution scale for the process, and $Q^2$, that is omitted in these formulae for ease of presentation.}
\begin{eqnarray}
\label{eq:3correlator}
W_{\Lambda,\Lambda'}^{\gamma^+} & = & \int \frac{dz^- }{(2\pi)} \, e^{i (X -\zeta/2) p^+ z^-}  \: 
\langle p'=p-\Delta ,\Lambda' | \overline{\psi} \left(0 \right) \, \gamma^+ \,  \psi\left( z^- \right)|p,\Lambda \rangle ,  \nonumber \\
& = & \frac{1}{2P^+} \left\{ H_q(X,\zeta,t) \bar{u}(p-\Delta,\Lambda') \gamma^+ u(p,\Lambda) + E_q(X,\zeta,t) \bar{u}(p-\Delta,\Lambda') \frac{\sigma^{i+} \Delta_i}{2M} u(p,\Lambda)  \right\}. 
\end{eqnarray}
Isolating the GPD $H$ for an unpolarized quark inside an unpolarized proton at skewness $\zeta$ equal to 0, in momentum space the $H_q(X, 0, t)$ is off-diagonal in momentum, 
\begin{eqnarray}
\label{eq:GPD3}
H_q(X, 0 ,t) & = &  \int d^2 {\bf k }_{T, in} \,   \phi^*(X,{\bf k}_{T, in} - {\bf \Delta}) \phi(X,{\bf k}_{T, in}),  
\end{eqnarray}
and with Fourier transformations of the vertex functions, 
\begin{subequations}
\label{eq:Fourier1}
\begin{eqnarray}
{\phi}(X,{\bf k}_{in}) & = & \frac{1}{(2 \pi)^2} \int d^2 {\bf z}_{in} \, e^{i {\bf k}_{in} \cdot {\bf z}_{in} } \, \tilde{ \phi}(X, {\bf z}_{in}), \\  
{\phi}^*(X,{\bf k}_{out}) & = & \frac{1}{(2 \pi)^2} \int d^2 {\bf z}_{out} \, e^{-i {\bf k}_{out} \cdot {\bf z}_{out} } \, \tilde{ \phi}^*(X, {\bf z}_{out}) . 
\end{eqnarray}
\end{subequations}
the relationship between the GPD and the IPPDF emerges: 
\begin{subequations}
\label{eq:H_to_rho}
\begin{eqnarray}
H_q(X, 0, t) &  = & \int d^2 {\bf b} \, e^{ i {\bf b}  \cdot \bf{ \Delta} } \, 
\tilde{ \phi}^*\left(X,{\bf b} \right) \, \tilde{ \phi}\left(X, {\bf b}\right) =  \int d^2 {\bf b} \, e^{ i {\bf b}  \cdot \bf{ \Delta} } \,  \rho_q(X,{\bf b}) \\
\rho_{q}(X, \bf{b}) &=& \int \frac{d^{2} \bf{\Delta}}{(2 \pi)^{2}} e^{- i \bf{b} \cdot \bf{\Delta}} H_{q}(X, 0, t), 
\end{eqnarray}
\end{subequations}
where we have defined the transverse space one-body diagonal density distribution. The index $q$ refers to the quark flavor. A similar distribution is ibtained for gluons. 

\begin{figure*}[h]
    \centering
    \includegraphics[width=12cm]{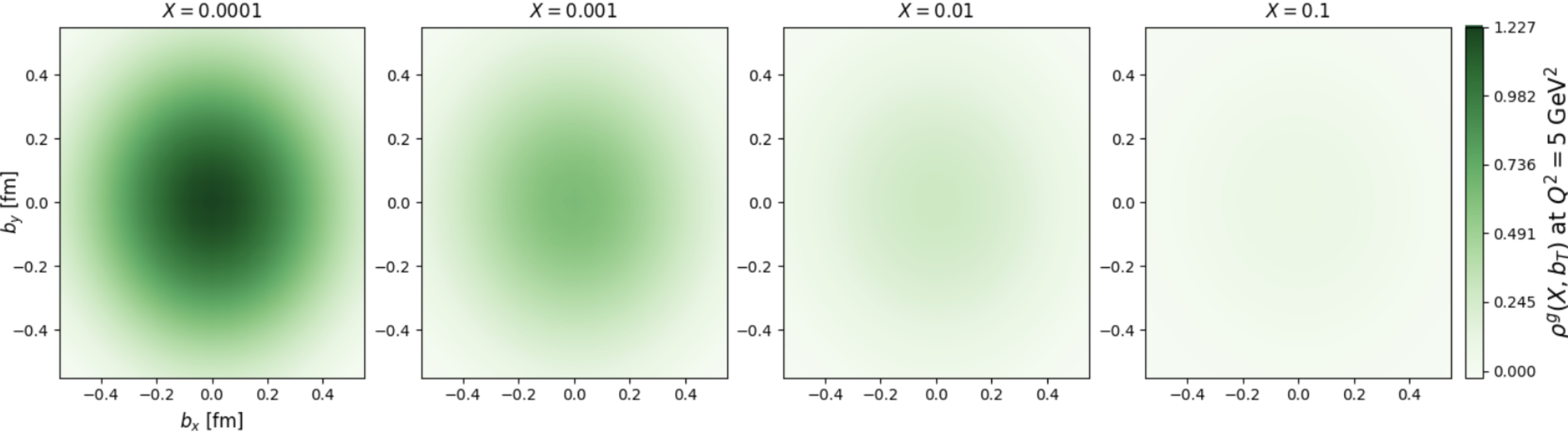}
    \includegraphics[width=8cm]{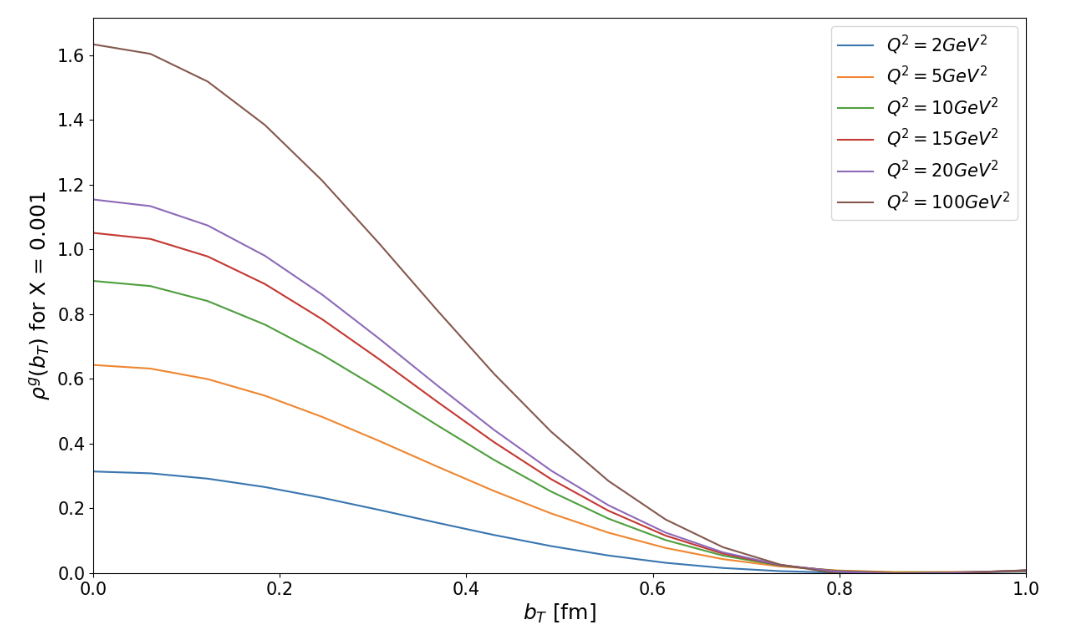}
    \caption{The Fourier transform $\rho_{g}(x,b)$ of the unpolarized gluon GPD $H_{g}$ calculated as a function of $x$ and $Q^{2}$.}
    \label{fig: spatial_g}
\end{figure*}

In Figure  \ref{fig: spatial_g} we show an example of a one-body density distribution, $\rho_{g}(X,b)$, evaluated for the gluon GPD parametrization from Refs.\cite{Kriesten:2021sqc,Goldstein:2010gu}, where $H_g$ was constrained by its first moment in $X$ evaluated in lattice QCD \cite{Shanahan:2018pib,Pefkou:2021fni}.  Notice that in this case, the gluon distribution depends rather strongly on the scale for the process, $Q^2$.

\begin{figure*}[h]
    \centering
    \includegraphics[width=7cm]{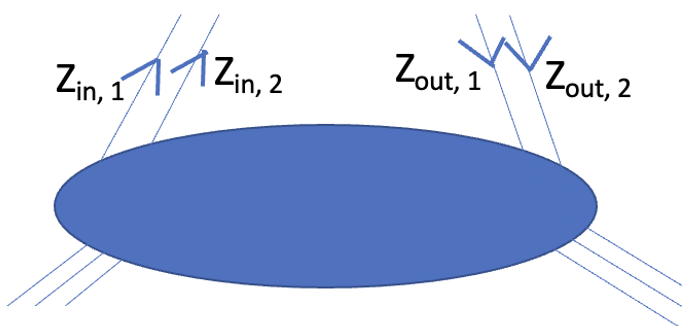}
    \caption{Two-body correlation function}
    \label{fig: two_body_corr}
\end{figure*}

Although the one-body density distributions already give rich spatial information about the proton's internal structure, as we show in Section \ref{sec: numerical}, in order to study the dynamics of the proton's constituents, we must move from a single-body to a two-body picture, studying now double-parton correlations. We thus introduce the two-body correlation function, as in Figure \ref{fig: two_body_corr}, defined through the following bilinear expression, 
\begin{eqnarray}
\label{eq:twocorr}
W_{\Lambda,\Lambda'}^{\Gamma} & = & \int \frac{dz_{1, in}^{-}d{\bf z}_{1, T, in}}{(2\pi)^{3}} \frac{dz_{2, in}^{-}d{\bf z}_{2, T, in}}{(2\pi)^{3}} \int \frac{dz_{1, out}^{-} d{\bf z}_{1, T, out}}{(2\pi)^{3}} \frac{dz_{2 out}^{-} d{\bf z}_{2, T, out}}{(2\pi)^{3}}\,  \nonumber \\
& \times &  e^{i  (k_{1, in} z_{1, in} + k_{2, in} z_{2, in})} \, e^{-i  (k_{1, out} z_{1, out} +k_{2, out} z_{2, out})}  \, \\ & \times & \langle p',\Lambda' | \overline{\psi} (z_{1, out}) \, \Gamma \psi(z_{1, in}) \, \overline{\psi} (z_{2, out}) \, \Gamma \psi(z_{2, in}) |p,\Lambda \rangle \Big | _{z_1^{+}= z_2^+ = 0}  \nonumber
\end{eqnarray}
This definition involving quark fields is similar to the double parton distribution function (DPDF) introduced in \cite{Diehl:2011yj}, but with the initial and final proton states being different, ($p' \neq p$). 
\begin{figure*}[h]
    \centering
    \includegraphics[width = 7cm]{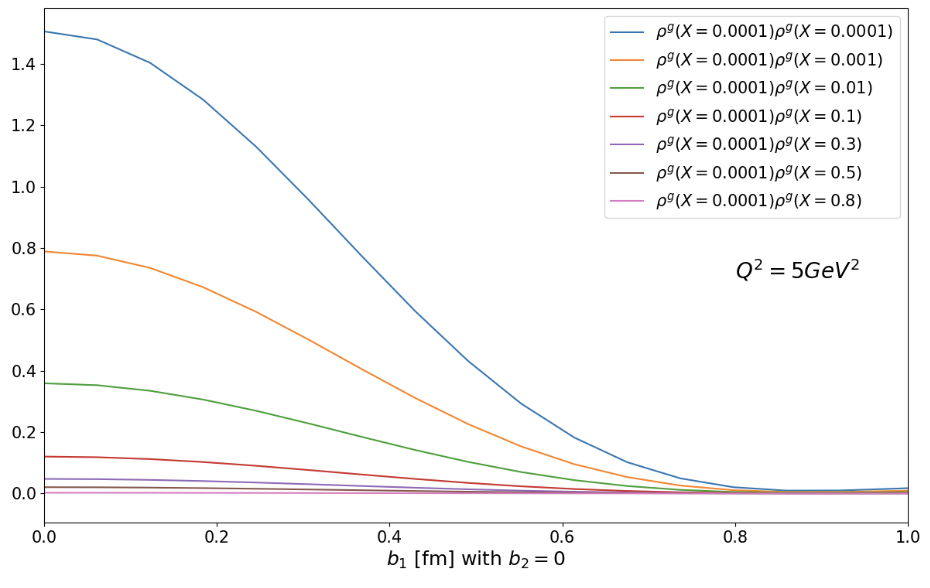}
    \includegraphics[width = 12cm]{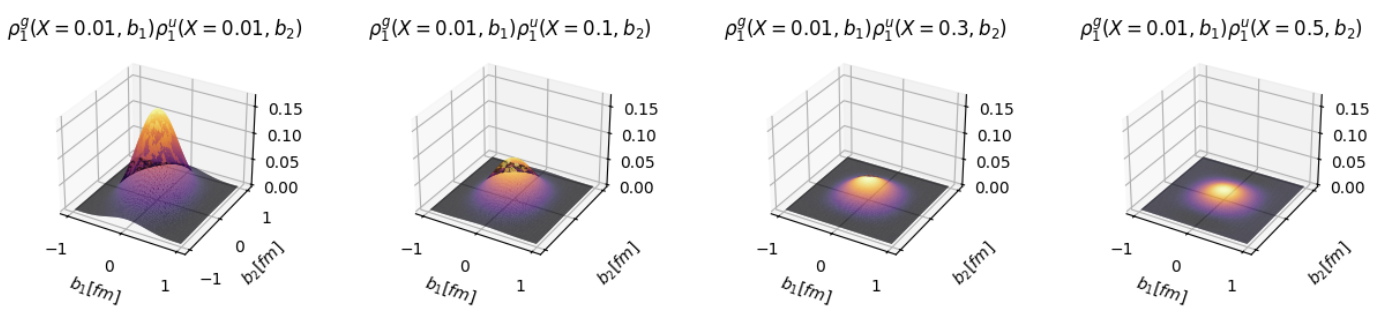} 
    \includegraphics[width = 12cm]{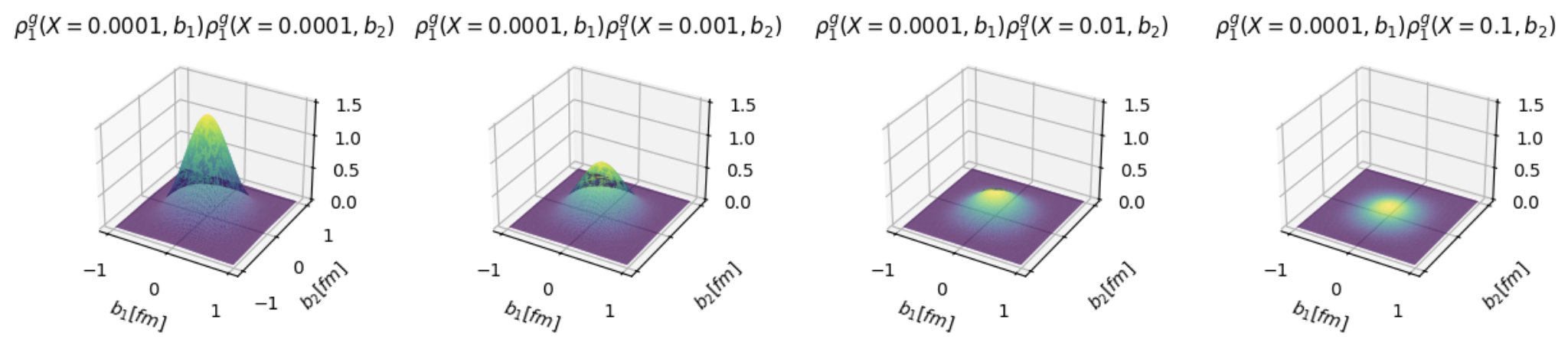}
    \caption{Visualizing the two-body density as a function of $b_{1}$ and $b_2$ (Eq.\ref{eq:rho2qg}), for different values of $x_2$, while keeping $x_1$ fixed.  }
    \label{fig: two_body_dens}
\end{figure*}
Proceeding similarly to the one-body case, we can now form the following combinations of the two different quark fields locations, $z_{i, in}$ and $z_{i, out}$, $(i=1,2)$, respectively  given by,
\begin{eqnarray}
\displaystyle\left\{\begin{array}{cc}
   z_i  & = z_{i, in} - z_{i, out} \\
   & \\
   b_i & = \displaystyle\frac{1}{2}(z_{i, in}+z_{i, out})  
\end{array} \right. \quad\quad
\Rightarrow 
\quad 
\displaystyle\left\{\begin{array}{cc}
   z_{i, in}  & = b_i + \displaystyle\frac{z_i}{2} \\
   &  \\
   z_{out} & = b_i - \displaystyle\frac{z_i}{2}  
\end{array} \right.  ,
\end{eqnarray}
as well as the conjugate momenta to $b_i$ and $z_i$, 
\begin{eqnarray}
\displaystyle\left\{\begin{array}{cc}
   \Delta_i  & = k_{i, in} - k_{i, out} \quad\quad \leftrightarrow \quad \quad b_i \\
   & \\
   k_i & = \displaystyle\frac{1}{2}(k_{i, in}+k_{i, out})  \quad\quad \leftrightarrow \quad \quad z_i 
   \label{eqn: out_in}
\end{array} \right. 
\end{eqnarray}
Following a procedure similar to the one-body case, we derived a formulation for the Fourier transform of the four-point correlation function \eqref{eq:twocorr}, showing that it can be cast in a two-body density form, $\rho^{qq}_2(x_1, {\bf b}_1; x_2,  {\bf b}_2)$,
describing the  probability of simultaneously finding quark $1$ carrying a momentum fraction $x_{1}$ at location ${\bf b}_1$ and quark  $2$ carrying a momentum fraction $x_{2}$ at location ${\bf b}_2$ with respect to the center of momentum of the system.  In the absence of correlations in the relative motion of the two quarks, the two-body transverse spatial distribution is described by,
\begin{eqnarray}
\label{eq:rho2qg}
 \rho^{qq}_2(X_1, {\bf b}_1; X_2,  {\bf b}_2) = \rho_q(X_{1}, {\bf b}_1) \rho_q(X_{2}, {\bf b}_2) 
\end{eqnarray}
where $\rho$ is the diagonal one-body density defined in Eq.\eqref{eq:H_to_rho}. 
Distributions with parton configurations other than $qq$, {\it e.g.} $qg$, or $gg$, can be written in a similar way, starting from the appropriate two-body correlation function definition.

A visualization of the two-body density distribution,  is provided in Fig. \ref{fig: two_body_dens} where in the upper panel we show a 2D rendition of the gluon-gluon distribution obtained by plotting the quantity, $\rho_2^{gg}(X_1,{\bf b}_1; X_2, {\bf b}_2)$ at the scale, $Q^2=5$ GeV $^2$, for $X_1= 10^{-3}$, and fixed ${\bf b}_2$, varying $X_2$ and ${\bf b}_1$; in the lower panels, we show 3D versions obtained at $x_1=0.01$ (top) and $X_1=10^{-3}$, plotted in the ${\bf b}_1$, ${\bf b}_2$ for various values of $X_2$.

\section{Numerical results/Observables} 
As described in the previous section, since the Fourier transform of the GPD $H$ for zero skewness gives the IPPDF, we obtain the single-particle spatial density of the parton's charge inside the proton (see {\it e.g.} Fig. \ref{fig: spatial_g}  for the gluon distribution). From such single-body densities, we can compute the average impact parameter squared, $\langle {\bf b}^{2} \rangle$, or the average transverse radius, as,
\begin{equation}
\langle {\bf b}_{q,g}^2 (X) \rangle   =  \frac{1}{{\cal N}_{q, g}} \displaystyle\int d^2 {\bf b}  \, {\bf b}^2 \rho_{q,g}(X,{\bf b}) ,
, g\end{equation}
where,
\begin{equation}
    {\cal N}_{q,g} = \displaystyle\int d^2 {\bf b}  \, \rho_{q,g}(X,{\bf b}) = {H_{q,g}(X,0,0)}  .
\end{equation}
\label{sec: numerical}
\begin{figure*}[h]
    \centering
    \includegraphics[width = 7cm]{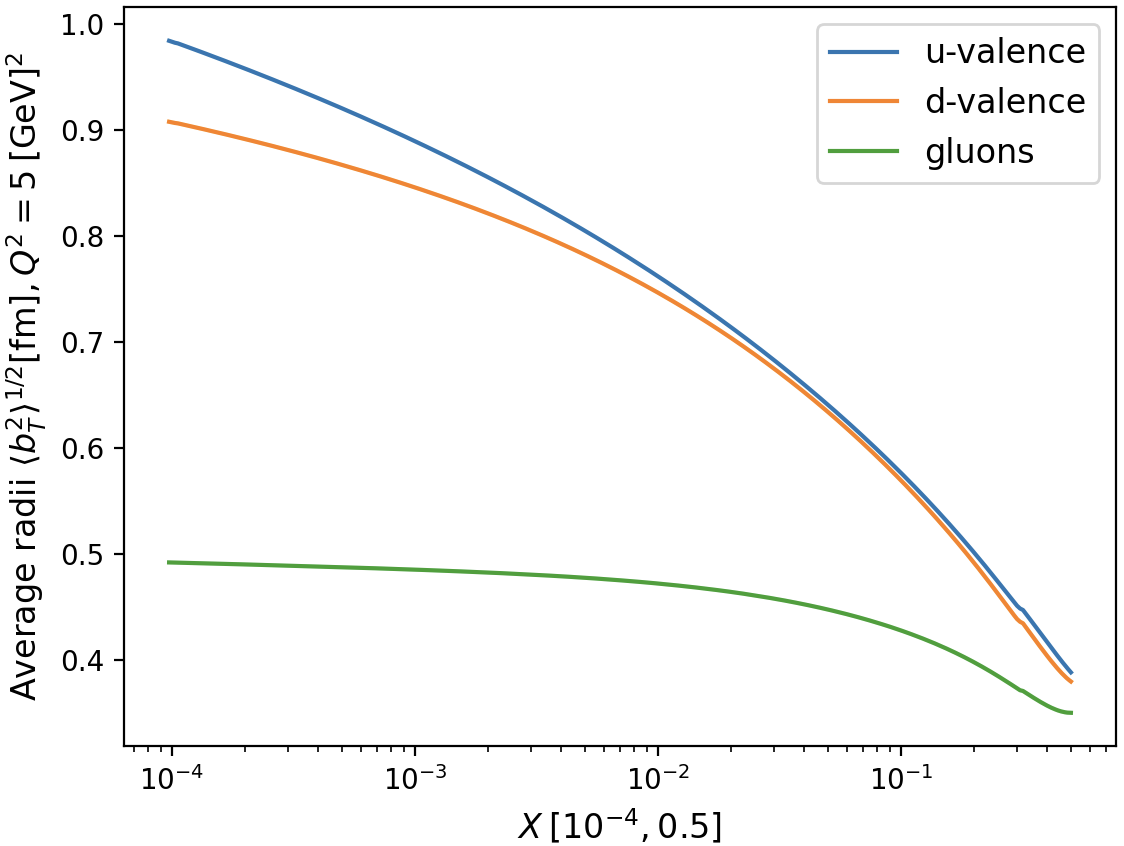}
    \includegraphics[width = 7cm]{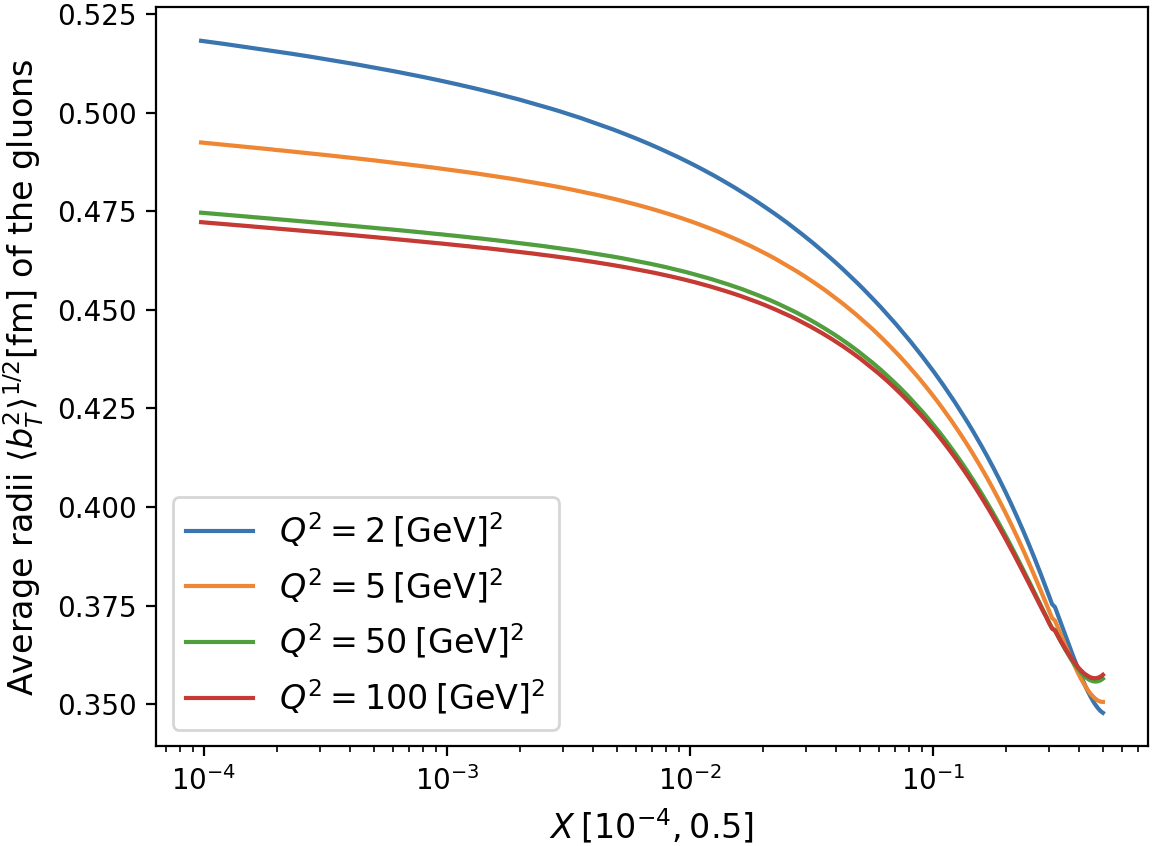}
\caption{The average radii for $u, d$ (valence) and $g$ (gluon) distributions (left) provide a quantitative method for extracting spatial information through even the one-body density. We can also study the $Q^{2}$-dependence of these average radii through perturbative evolution, as shown for the gluons (right).}
\label{fig:radii}
\end{figure*}
This normalization encapsulates the baryon number, putting the partonic distributions on equal footing to make a faithful comparison. 
The average radii are shown in Figure \ref{fig:radii} where on the {\it lhs} we plotted the valence $u$ and we $d$ quark geometric radii compared to the gluon one, at $Q^2$= 5 GeV$^2$; on the {\it rhs} the behavior of the gluon radius with $Q^2$ is shown. All calculations were performed using the parametrization from Refs.\cite{Kriesten:2021sqc,Goldstein:2010gu}. One can see that the gluon radius is much smaller than the quark one, consistent with the geometric representation of baryon junctions, as proposed in Ref.\cite{Kharzeev_1996} and recently investigated in, $\it{e.g.}$, Ref.\cite{PhysRevD.108.054002}. Furthermore, we find that, while the radii for all $q$ and $g$ components show a small variation with $Q^2$ at larger values of $X > 0.01$, at small $X$, this dependence is substantial, and it should, therefore be more easily detectable in experiment.

From the vectors locating the partons 1 and 2 positions in the transverse plane, ${\bf b}_1$ and ${\bf b}_2$, respectively, we define the relative distance between them, ${\bf r}$, and the distance of the center of mass of the two partons from the origin, ${\bf R}_{12}$,
\begin{eqnarray}
    {\bf r} & = & {\bf b}_1 - {\bf b}_2 \\
    {\bf R}_{12} & = & \frac{{\bf b}_1 + {\bf b}_2}{2}
\end{eqnarray}
Defining the root mean squared of their expectation values, one obtains the square average relative distance and average center of mass of the two partons, 
\begin{eqnarray}
\label{eq:distance}
 \langle {\bf r}^2  \rangle(X_1,X_2) &  =  & \frac{1}{{\cal N } }  \int \int d^2{\bf r} \, d^2{\bf R}_{12}  \, 
\left| {\bf r}^2 \right| \,\,
\rho_{2} \left( X_1,  {\bf R}_{12} + \frac{{\bf r}}{2}; X_2, {\bf R}_{12} - \frac{{\bf r}}{2}  \right) 
\\
\label{eq:R12}
 \langle {\bf R}_{12}^2 \rangle (X_1,X_2) &  = &  \frac{1}{{\cal N } }  \int \int d^2{\bf r} \, r^2 {\bf R}_{12}  \, 
\left| {\bf R}_{12}^2 \right| \,\, \rho_2 \left( X_1, {\bf R}_{12} + \frac{{\bf r}}{2}; X_2, {\bf R}_{12} - \frac{{\bf r}}{2}  \right)  \\
{\cal N } & = & \int \int d^2{\bf r} \, r^2 {\bf R}_{12}  \,\, 
\rho_2 \left( X_1, {\bf R}_{12} + \frac{{\bf r}}{2}; X_2, {\bf R}_{12} - \frac{{\bf r}}{2}  \right) 
\end{eqnarray}
\begin{figure*}[h]
    \centering
    \includegraphics[width = 7cm]{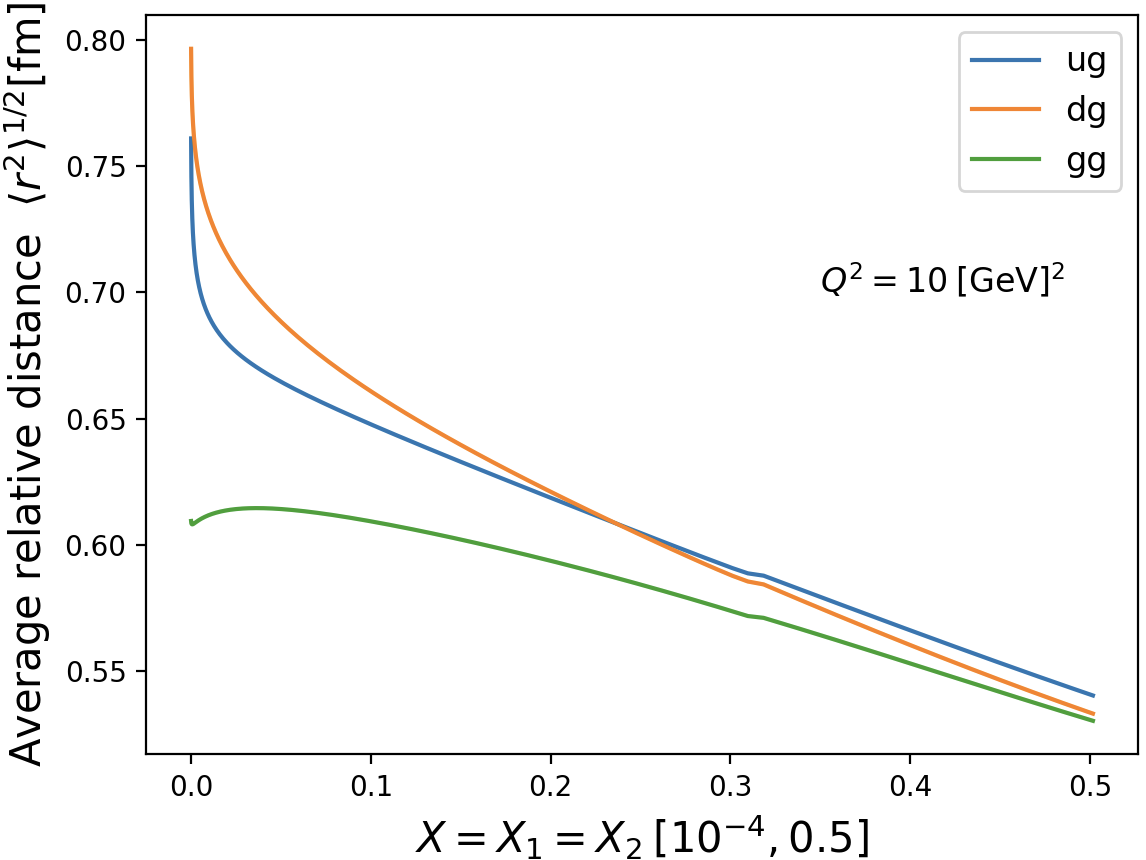}
    \caption{Average relative distance between the $u$ quark and gluon, the $d$ quark and gluon, and two gluon fields inside the proton plotted vs. $X=X_1=X_2$, at $Q^2$ = 10 GeV$^2$.}
    \label{fig: rel_dist_overlap_1}
\end{figure*}
The numerical results obtained using our parametrization for the average relative distance, given in Fig. \ref{fig: rel_dist_overlap_1}, for the specific case of $X_1=X_2$, correlate with the results for the single parton distributions, in that the distance between a quark of flavor $u,d$ and a gluon field is always larger than the distance of two gluon fields, revealing that gluonic configurations tend to be more compact.

Through the two-body densities we can now address quantitatively the question of whether the $u,$ quarks and gluons will overlap forming localized ``hot spots," as suggested in an event-by-event analysis in Refs. \cite{Mantysaari:2020axf,Mantysaari:2020lhf}, or whether the gluon field surrounds uniformly the valence quarks. 
The geometric average two-parton overlap probability in the transverse plane can be defined as,
\begin{eqnarray}
   \langle A(X_1,X_2) \rangle & = & \frac{1}{{\cal N}} \int d^2 {\bf b}_1 d^2 {\bf b}_2 \, \rho_2^{ij}(X_1,{\bf b}_1; X_2, {\bf b}_2)  A(\mid {\bf b}_1 - {\bf b}_2 \mid ), \nonumber
\end{eqnarray}
where $A(r=\mid {\bf b}_1 - {\bf b}_2\mid )$ is taken as the geometric area overlap of the two azimuthally symmetric single parton distributions measured in units of an average radius $a =  (\langle b_1 \rangle^2 + \langle b_2 \rangle^2)^{1/2}/\sqrt{2}$, 
\begin{eqnarray}
%
    A(r) & = & a^2 \pi  - \frac{r -\delta}{2}  \left( \alpha_1 + \alpha_2 \right)  
    - a_1^2 \arctan{\frac{r-\delta}{ 2 \alpha_1} } - a_2^2 \arctan{\frac{r-\delta}{2 \alpha_2} }
\end{eqnarray}
with, 
\begin{eqnarray}
\alpha_i =   \sqrt{a_i^2- \frac{(r-\delta)^2}{4}} , \quad \quad i=1,2
\end{eqnarray}
and $\delta= a_1-a_2$. 

\begin{figure*}[h]
    \centering
    \includegraphics[width = 6.5cm]{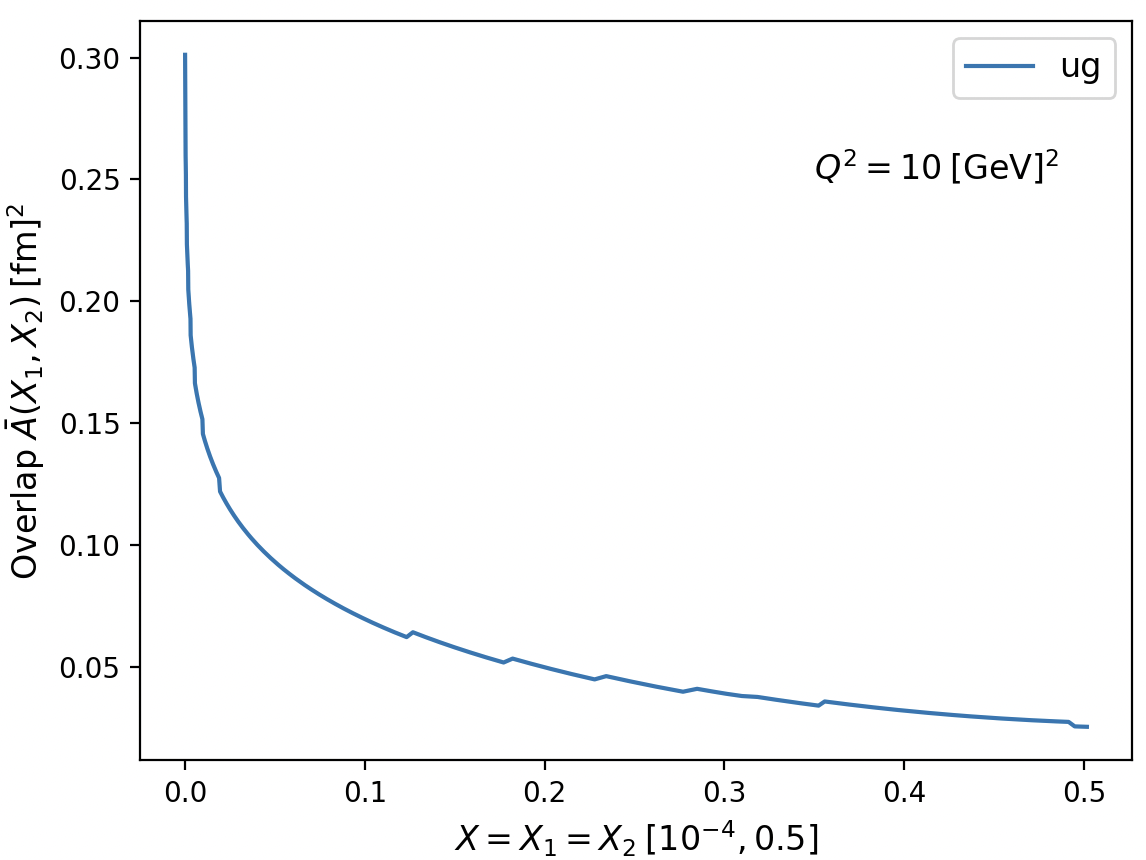}
    \caption{The overlap probability of the $u$ quark and gluon distributions in the proton plotted vs $X=X_1=X_2$ for $Q^2 = 10 $ GeV$^2$.}
    \label{fig: rel_dist_overlap_2}
\end{figure*}

\noindent 
Performing a change of variables, the average two-parton overlap probability is given as,
\begin{eqnarray}
\label{eq:overlap}
\langle {A}(X_1, X_2) \rangle  & = & \frac{1}{{\cal N } }  \int \int d^2{\bf r} \, d^2{\bf R}_{12}  \, 
A(r) \, \rho_2^{qg} \left( X_1, {\bf R}_{12} + \frac{{\bf r}}{2}; X_2,  {\bf R}_{12} - \frac{{\bf r}}{2}  \right) . 
\end{eqnarray}
Numerical results for the overlap probability are given in Figure \ref{fig: rel_dist_overlap_2} for the $u$ quark and gluon case. The figure shows $A(X_1,X_2)$ , Eq.\eqref{eq:overlap}, plotted vs. $X$, with  $X= X_1=X_2$, and $Q^2 = 10 $ GeV$^2$, divided by the maximum overlap surface. Putting together results from Figs.\ref{fig:radii}, \ref{fig: rel_dist_overlap_1}, and \ref{fig: rel_dist_overlap_2}, we can see that while all parton types tend to be more broadly distributed at low $X$ (Fig.\ref{fig:radii}), thus also spanning broadly distributed relative distances, their overlap probability also increase (Fig.\ref{fig: rel_dist_overlap_2}). 
This picture complements the recent BNL approach \cite{Mantysaari:2019jhh,Mantysaari:2020axf,Mantysaari:2020lhf,Dumitru:2021tvw},  describing geometrical fluctuations, or event-by-event fluctuations in the proton wave function where the three constituent quarks emit small$-x$ gluons around them, which form  ``hot-spots" at random locations in the transverse plane. 

\section{Conclusions}
\label{sec:conclusions}
We presented results on our study of two-body parton distributions as a means of obtaining information on the relative positions of quarks and gluons inside the proton. 
While one-body distributions are limited to capturing average static properties of the proton's internal structure, such as the radius covered by various parton distributions, the proton is a complex, multi-body environment, and it is, therefore, more properly imaged through multi-body distributions. As a first step towards the general problem of imaging this  multi-body system, and to provide tools to address multiparton correlations beyond the one-body formalism, we studied the geometric correlations between the $u$ and $d$ quark distributions and gluons.

In Refs.\cite{Mantysaari:2020axf,Mantysaari:2020lhf} a  connection was provided between the two approaches, ``GPDs" on one side, and ``geometrical fluctuations" on the other. In particular, the dipole-based description of {\it coherent} diffractive $J/\psi$  production off a proton was connected to the correlation functions for GPDs, within the collinear framework of QCD. 
However, the more interesting channel of {\it incoherent} diffraction, described in terms of  double dipole diffraction 
and leading to small$-x$ gluon fluctuations, or event-by-event, color density geometry driven 
fluctuations has remained a puzzle in that it could not be rendered, so far, within standard QCD approaches. 
By introducing two-body correlations functions, we provide a framework to investigate these scenarios and to allow us to study relative spatial configurations of quarks and gluons inside the proton and the atomic nuclei.

\acknowledgments
This research is funded by DOE grant DE-SC0016286.

\bibliographystyle{unsrt}
\bibliography{BIB}

\end{document}